\documentclass[10pt]{article}

\usepackage{arxiv}

\newcommand{\copyrightnotice}[1]{{%
  \renewcommand{\thefootnote}{}
  \footnotetext[0]{#1}%
}}

\usepackage{multicol}
\usepackage{fullpage,graphicx,psfrag,amsmath,amsfonts,verbatim}
\usepackage{amsmath,amsfonts,amssymb}
\usepackage{latexsym, amscd, amsfonts, eucal, mathrsfs, amsmath, amssymb, amsthm, xypic,xr, stmaryrd, color, enumerate, mathtools}
\usepackage{amsmath,amsfonts,amssymb,soul}
\usepackage[utf8]{inputenc}

\usepackage{natbib}
\usepackage{hyperref}

\MakeRobust{\ref}

\makeatletter
\newcommand{\labeltext}[2]{%
  \@bsphack
  \csname phantomsection\endcsname 
  \def\@currentlabel{#1}{\label{#2}}%
  \@esphack
}
\makeatother

\newcommand{\multilinecomment}[1]{}

\DeclareSymbolFont{bbold}{U}{bbold}{m}{n}
\DeclareSymbolFontAlphabet{\mathbbold}{bbold}

\DeclareMathOperator*{\argmax}{arg\,max}

\usepackage{fancyhdr}

\newcommand{\stratnum}{n}
\newcommand{\propnegative}{p}
\newcommand{\respi}{r_i}
\newcommand{\respj}{r_j}

\newcount\Comments  
\definecolor{darkgreen}{rgb}{0,0.5,0}
\newcommand{\kibitz}[2]{\ifnum\Comments=1\textcolor{#1}{#2}\fi}

\newcommand{\game}{G}
\newcommand{\subjgame}{\game(\alpha_i,\alpha_j)}

\newcommand{\oligone}{G_{\kappa_1}}
\newcommand{\oligtwo}{G_{\kappa_2}}

\newcommand{\nash}{\text{NE}}
\newcommand{\br}{\text{BR}}
\newcommand{\behavioral}{\mathcal{B}}

\newcommand{\rational}{\mathcal{R}}

\newcommand{\handshake}{\mathcal{H}}

\newcommand{\actspace}{\mathcal{A}}
\newcommand{\stratspace}{\mathcal{S}}
\newcommand{\complexfunc}{c}
\newcommand{\multicomplex}{c_K}

\newcommand{\multbehavioral}{\mathcal{B}_K}

\newcommand{\multrational}{\mathcal{R}_K}

\newcommand{\act}{a}

\newcommand{\fit}{f}
\newcommand{\pay}{u}
\newcommand{\nashpay}{\pay_N}

\usepackage{makecell}
\usepackage{algorithm}
\usepackage[noend]{algpseudocode}

\usepackage{tikz}

\usetikzlibrary{bayesnet, arrows.meta,snakes}
\usetikzlibrary{matrix, fit}
\usetikzlibrary{decorations.markings}
\tikzset{
    myarrow/.style={
        -Stealth,
        shorten >=8pt,
        shorten <=8pt,
    },
    mymatrix/.style={
        matrix of math nodes
    },
}

\newtheorem{theorem}{Theorem}
\newtheorem{definition}{Definition}

\newtheorem{proposition}[theorem]{Proposition}

\theoremstyle{definition}

\usepackage{mathtools}

\title{Evolutionary Stability of
Other-Regarding
Preferences Under Complexity Costs}
\author{ Anthony DiGiovanni \\
	Center on Long-Term Risk\\
	\And
	Nicolas Mac\'e \\
	Center on Long-Term Risk\\
	\And 
	Jesse Clifton \\
	Center on Long-Term Risk
}
\date{}

\begin{document}

\maketitle
\copyrightnotice{\!\!\!\!\!\!\!\!\!\!\!Presented at the Learning, Evolution and Games 2022 workshop.}

\begin{multicols}{2}

\begin{abstract}
The evolution of
preferences that account for other agents' fitness, or
\textit{other-regarding preferences},
has been modeled with the ``indirect
approach'' to evolutionary
game theory. 
Under the indirect evolutionary approach, 
agents make decisions by optimizing a subjective
utility function. Evolution may select
for subjective preferences that 
differ from the fitness function, and in
particular, subjective preferences for
increasing or reducing other agents' fitness.
However, 
indirect evolutionary models typically artificially restrict the space of strategies that agents might use
(assuming that agents always play a Nash equilibrium under their subjective preferences), 
and dropping this restriction can undermine the finding that other-regarding preferences are selected for. Can the indirect evolutionary approach still be used to explain the apparent existence of other-regarding preferences, like altruism, in humans? We argue that it can, by accounting for the costs associated with the complexity of strategies, giving (to our knowledge) the first account of the relationship between strategy complexity and the evolution of preferences.
Our model formalizes the intuition
that agents face tradeoffs between
the cognitive costs of strategies
and how well they interpolate across
contexts.
For a single game, these complexity costs lead to selection
for a simple fixed-action strategy, 
but across games, 
when there is
a sufficiently large
cost to
a strategy's number
of context-specific
parameters,
a strategy of maximizing subjective (other-regarding)
utility
is stable again.
Overall, our analysis provides a more nuanced picture
of when other-regarding preferences will evolve.

\end{abstract}

\section{Introduction}

Under what conditions do agents
evolve to maximize a subjective utility function
other than their evolutionary fitness?
In particular, when is there selection
for \textit{other-regarding preferences} 
\citep{sour_grapes,social_choice}
such as
altruism (intrinsically valuing
improvements in other
agents' fitness) 
or spite (intrinsically valuing
reductions in other agents' fitness)? 
These questions have been previously studied under the ``indirect approach'' to evolutionary game theory \citep{indirect}.
Consider a game whose payoffs determine
the players' fitness in an evolutionary
process, called a base game.
The indirect evolutionary approach supposes that
selection occurs on agents' subjective preferences
(hereafter, ``preferences'')
represented
as utility functions,
and agents rationally play the base game by 
optimizing their subjective utility functions.
When assessing the evolutionary stability of strategies
in the indirect approach,
a player's utility function defines their
strategy.
This is in contrast to the classical ``direct''
approach where actions in the base game themselves are selected.

This indirect approach has been applied in attempts to
explain
altruism in organisms,
especially in contexts where other explanations such as kin selection
and reciprocity are inadequate \citep{bester1998altruism,trustworthy_oneshot,outgroupspite}.
In a simple model of an interaction where two agents' actions
have positive externalities for each other --- i.e., increasing one's
action (represented as a real number) increases the other's
payoff
--- \citet{bester1998altruism}
find that altruistic preferences 
are evolutionarily stable.
\citet{bolle2000altruism} and \citet{possajennikov2000evolutionary} extended this model to also
explain the stability of spiteful preferences
in interactions with negative externalities.
These other-regarding preferences
are selected because they are known to other
agents, and thus
credibly signal an agent's commitment
to certain behavior, given other
agents' preferences \citep{blush,commitment}.

However, these models have two key limitations: 
\begin{enumerate}
    \item They assume that agents always play a best
    response given their 
    preferences and
    beliefs about the other player’s 
    preferences.
    This precludes agents who commit to following a certain action 
    regardless of their beliefs about the other player.
    This is important because, as we will show, when such commitments are allowed,
    a subjective utility-maximizing strategy with 
    other-regarding preferences
    is no longer the unique evolutionarily stable strategy.
    \item 
    They restrict the space of preferences in a way that
    prevents the use of strategies
    capable of invading populations of inefficient strategies,
    called the ``secret handshake'' in previous work \citep{handshake}.
    As \citet{evopref} show,
    when the space of preferences is expanded
    to include all possible utility functions,
    evolutionarily
    stable strategies in an indirect evolutionary model 
    must be efficient.
    This is
    because any population of inefficient strategies can 
    be invaded by mutants who mimic the behavior of the inefficient
    strategy,
    and play an efficient 
    action
    against other mutants.
\end{enumerate}

These two modifications to the original 
indirect evolutionary models undermine those models’ conclusions
that other-regarding preferences
can be evolutionarily stable, including
preferences that lead to inefficient behavior.
However, an important feature of the kinds of strategies
described in (1) and (2) is that they differ from subjective
utility
maximization in 
their \textit{complexity costs},
i.e.,
the costs an agent must pay to learn and execute
strategies \citep{richer_evo}.
These costs may play a critical role in evolution;
for instance, 
the tradeoff between the problem-solving
benefits and energetic costs of larger
brains
may explain 
variation in brain size among primates,
and in animal behavior in contests
\citep{brains,contest}.
Previous literature has studied how complexity costs affect the evolutionary
stability of strategies \citep{automata_original,automata_base_ne,ipd_automata}.
The costs of strategy complexity
accumulate over the diverse set of environments and
interactions an agent faces in its lifetime \citep{evo_generalization}.
Thus, instead of using many different
strategies that are each simple in isolation,
it can be less expensive overall for an
agent to use a sophisticated strategy that 
interpolates
well across interactions \citep{evo_strategic_soph, waysofaltruism}.
We will argue that the complexity costs of
applying individualized heuristics to
each new interaction may
be sufficiently high
that evolution selects
for ``rational'' agents,
which consistently
optimize some (other-regarding) utility function.

Our key contribution is a revised account 
of the evolution of other-regarding preferences,
based on a novel framework accounting for
the fitness costs that strategies incur due to their complexity
in \textit{multiple} strategic contexts.
While existing indirect evolutionary models are inadequate because they
artificially restrict the space of strategies,
we show that their predictions can be
recovered by accounting for how subjective utility-maximizing strategies optimally 
trade off complexity within and across decision contexts.
In particular:
\begin{itemize}
    \item
    We characterize the Nash equilibria (and
    stability thereof) of the space of subjective utility-maximizing strategies from \citet{possajennikov2000evolutionary}
    augmented with strategies that commit to 
    a certain action (``behavioral strategies''),
    in a general class of symmetric two-player
    games.
    In this expanded space,
    rational strategies
    with other-regarding preferences that are
    evolutionarily stable against other rational
    strategies,
    as in \citet{bester1998altruism} and \citet{possajennikov2000evolutionary},
    are no longer the unique evolutionarily stable
    strategies.
    This result motivates the search for an alternative
    explanation of the evolution of 
    other-regarding preferences.
    \item 
    While previous work has shown how finite computational
    costs of strategies in repeated games significantly alter
    the set of stable strategies,
    we present two results
    illustrating a tradeoff between within-game and across-game complexity costs:
    (1) 
    Suppose that
    rational strategies are more costly 
    in a single interaction than behavioral strategies,
    given 
    the greater
    energetic costs associated with their complex cognition
    \citep{Conlisk1980Sep,abreu2003evolutionary}.
    Then in an \textit{individual} complexity-penalized game, the multiplicity of
    neutrally stable
    strategies, including rational strategies with
    other-regarding preferences, is
    replaced with a unique evolutionarily stable strategy, the Nash equilibrium of the base game.
    (2) When agents play multiple games, a sufficiently large
    penalty on the number of game-specific parameters used by 
    a strategy reproduces the results of \citet{bester1998altruism} and \citet{possajennikov2000evolutionary}
    --- in numerical experiments, 
    the population converges (under a particular
    evolutionary dynamic) to a rational
    strategy with other-regarding preferences.
    Our experiments also explore how the size of
    the
    penalty on game-specific parameters necessary for
    other-regarding preferences to evolve, and the
    strength of altruism or spite that evolves,
    depend on the distribution of games.
    \item We argue that accounting
    for complexity costs blocks the secret handshake argument:
    Mutant strategies that both mimic
    an inefficient action and play
    an efficient action against themselves are
    more complex than behavioral strategies,
    and thus cannot invade a strategy
    that always plays the Nash equilibrium of the base game.
\end{itemize}

\section{Related Work}

\paragraph{Indirect evolutionary approach}
Like \citet{bester1998altruism}, \citet{bolle2000altruism}, and \citet{possajennikov2000evolutionary},
we model rational players as
playing Nash equilibria with respect
to 
utility functions
given by their own fitness
plus a (possibly negative) multiple of their opponent's fitness.
\citet{Heifetz03whatto} generalize this model to
utility functions
given by one's own fitness plus some
function called a \textit{disposition}.
They show that
dispositions are not eliminated by selection
in a wide variety of games.
Generalizing further to the space of all
possible utility functions in finite-action
games, \citet{evopref} show that any strategy
achieving an inefficient payoff against itself ---
including the kinds of strategies with
other-regarding preferences predicted by
\citet{possajennikov2000evolutionary} ---
is not evolutionarily stable.
We will argue, however,
that the invader strategies that
make efficiency necessary for stability
are 
more complex than behavioral or rational
strategies,
and thus when 
complexity 
costs are accounted for in an 
agent's fitness,
a stable strategy can lead
to inefficiency.
\citet{incomplete_info_pref_evo} and \citet{ne_evo_prefs}
note that in order for
utility functions
to evolve
such that players with those
utility functions
do not play the
base game Nash equilibrium,
players must 
have information about
each other's
utility functions.
We assume
utility functions are known,
and briefly discuss how players can learn
each other's
utility functions
over repeated
interactions,
but acknowledge that this is a substantive assumption
since players often have incentives
to send deceptive signals of their 
utility functions
\citep{darwinnash}.
Finally, \citet{dynamic_evo} and \citet{hamilton} generalize
\citet{possajennikov2000evolutionary}'s finding that altruism
or spite can be evolutionarily stable in a certain game depending
on whether it features positive
or negative externalities.
They show that in a general class of
games,
selection for altruism versus spite
is determined (partly) by whether
the base game has strategic complements or substitutes,
i.e., whether increasing one player's input increases or decreases
the marginal value of input for another player.
The patterns of selection of altruism
or spite based on multiple games that we illustrate with \citet{bester1998altruism}'s game,
therefore, might hold for a variety of games.

\paragraph{Games with complexity costs} \citet{automata_original} 
characterizes Nash equilibria in repeated
games under computational costs.
He represents strategies in the repeated Prisoner's Dilemma
as finite-state automata
(sets of states determining the player's action
with rules for transitions between states).
Complexity costs are lexicographic:
an automaton achieving a strictly higher payoff is always preferred,
but when two automata achieve the same average
payoff, the automaton with fewer states
is preferred.
\citet{ipd_automata} show that although no
evolutionarily stable strategies exist in
the repeated Prisoner's Dilemma without complexity costs,
adding these lexicographic costs leads
to the existence of
some evolutionarily
stable strategies.
We similarly show that
in one-shot games,
when we account
for the greater complexity of ``rational'' strategies
relative to ``behavioral'' (fixed-action)
strategies,
a set of multiple
neutrally stable strategies
is replaced
with a unique evolutionarily stable strategy.
Our distinction between the complexity
of rational and behavioral
strategies follows that of \citet{abreu2003evolutionary},
who show that under an arbitrarily small cost of
the complexity of rationality,
behavioral strategies are evolutionarily
stable in a bargaining game.
If automata are also penalized based on the number of different
states each state can transition to,
the evolutionarily stable strategies
are restricted to the Nash equilibria of the
(non-repeated) base game \citep{automata_base_ne}.
We find an analogous result in 
one-shot games
with a different complexity metric.
Lastly, \citet{evo_repeated_complexity} find that
in the repeated Prisoner's Dilemma, increasing
non-lexicographic complexity costs
decreases the frequency
of cooperation
in
finite-population stochastic evolutionary simulations.
Similarly, in 
the multi-game setting,
we find numerically that
as
complexity costs on a strategy's number of
game-specific parameters
increase,
there are 
transitions between more or less
efficient stable strategies.

\paragraph{Coevolution of rationality and other-regarding preferences}
A key theme in our work is that selection
may favor the ability of rational agents,
which have other-regarding preferences and 
model
other players as optimizing their 
own
utility functions,
to solve a variety of strategic problems.
Building on \citet{util_function_evo}'s analogous results in single-agent problems,
\citet{evo_strategic_soph}
model
the
coevolution of
utility maximization and ability to attribute preferences to others.
Like us, they show that
after accounting for the advantages of 
interpolation
across strategic contexts, selection favors a rational strategy that learns and responds
to the preferences of its opponent,
as opposed to strategies that do not know how to respond to 
new games.
However, we study selection pressures towards
rationality in the context of evolution of preferences.
Further,
in our analysis, the advantage of rationality comes from
avoiding costs that non-rational strategies pay to adapt a
response to each separate game,
rather than from
non-rational strategies'
inability
to respond to new games.
\citet{darwinnash} model the evolution of both preferences
and the cognitive capacity necessary to
signal false preferences to
others.
Their argument for the efficiency of stable
strategies is vulnerable to the
counterargument
that we raise to
\citet{evopref}
above.
However, their results are similar to ours in that 
the set of stable strategies
is sensitive to whether the costs
of cognitive complexity are sufficiently high,
relative to the 
direct fitness benefits of complex cognition.
Like us, \citet{evo_generalization} model
the evolution of strategies that 
interpolate
across different contexts, but
their analysis is restricted to cooperation
in a certain class of games rather than evolution of other-regarding
preferences in general
(including uncooperative preferences
like spite).

\section{Preliminaries and Running Example}

We begin with definitions and notation 
and introduce
a well-studied game that
will illustrate principles of the indirect
evolutionary approach.

Let $\game$ be any symmetric two-player game
(called the base game)
between players $i$ and $j$,
with action space $\actspace$
and payoff functions $\pay_i, \pay_j: \actspace^2 \to \mathbb{R}$.
Players choose actions in the base game
as functions of strategies that are selected
in an evolutionary process.
Suppose 
players simultaneously play strategies
(elements of some abstract space $\Sigma$)
and observe each other's strategies,
then play $\game$
with
actions determined by
the pair of strategies.
Then, define the function $\respi: \Sigma^2 \to \actspace$, where
player $i$'s action in $\game$
given the players' strategies $\sigma_i, \sigma_j \in \Sigma$ is
$\respi(\sigma_i, \sigma_j)$.
In standard evolutionary analysis the fitness
of a strategy
equals
its payoff in $\game$,
thus we write
player $i$'s fitness from a strategy profile
as $\fit_i(\sigma_i,\sigma_j) = \pay_i(\respi(\sigma_i, \sigma_j),\respj(\sigma_j, \sigma_i))$.
(We distinguish fitness from payoffs because
once complexity costs are included, as in Section
\ref{sub:complexity}, this identity no longer holds.)
The following definitions classify a strategy based on
the robustness to mutations of a population purely consisting of that
strategy.

\begin{definition}
Relative to a fixed strategy space $\Sigma$ for $\game$,
a strategy $\sigma \in \Sigma$
is:
\begin{itemize}
    \item A \textbf{Nash equilibrium} if, for all $\sigma' \in \Sigma$,
$\fit_i(\sigma,\sigma) \geq \fit_i(\sigma',\sigma)$.
    \item A \textbf{neutrally stable strategy (NSS)} if (1) it
    is a Nash equilibrium, and (2\labeltext{2}{nss_two}) for all $\sigma'$ such that
    $\fit_i(\sigma',\sigma) = \fit_i(\sigma,\sigma)$, 
    $\fit_i(\sigma,\sigma') \geq \fit_i(\sigma',\sigma')$.
    \item An \textbf{evolutionarily stable strategy (ESS)} if
    it is an NSS and the inequality in \ref{nss_two} is always strict.
\end{itemize}
\end{definition}

The strict inequality $\fit_i(\sigma,\sigma') > \fit_i(\sigma',\sigma')$ in the definition
of ESS implies a stronger ``pull'' towards
an ESS in evolutionary dynamics (such as the replicator dynamic)
than towards an NSS:
If a rare mutant that enters a population consisting of an ESS
has the same fitness when paired with itself
as the ESS has against this mutant,
the mutant goes extinct under the replicator dynamic,
but this does not necessarily hold for an NSS
\citep{nss_props}.

Our running example is the following
symmetric two-player game, which we call the externality game \citep{bester1998altruism}.
Each player $i$ simultaneously chooses $a_i \in \mathbb{R}$,
and, for some $m > 0$ and $\kappa \in [-2, 0) \cup (0, 1)$, the players receive payoffs:
\begin{align*}
    \pay_i(a_1,a_2) &= a_i(\kappa a_j + m - a_i).
\end{align*}

Thus, $\kappa$ represents negative or positive externalities of
each player's action for the other's
payoff (when $\kappa < 0$ or $\kappa > 0$, respectively).
In the original model, players are assumed to have 
the following \textit{subjective utility functions}, for $\alpha_i \in \mathbb{R}$:
\begin{align*}
    V_i^{\alpha_i}(a_1,a_2) &= \pay_i(a_1,a_2) + \alpha_i \pay_j(a_1,a_2).
\end{align*}
Players behave rationally with respect to their subjective
utility functions,
and subjective utility functions are common
knowledge.
Thus the players play
the 
Nash equilibrium of the game
$\subjgame$
in which payoffs are
given by $V_i^{\alpha_i}, V_j^{\alpha_j}$,
denoted $\nash_i(\alpha_i,\alpha_j)$.
That is,
letting $\alpha_i$
represent player $i$'s strategy,
$r_i(\alpha_i,\alpha_j) = \nash_i(\alpha_i,\alpha_j)$.

\par A player with $\alpha_i > 0$ (respectively, $\alpha_i < 0$)
has subjective
utility increasing (decreasing) with
the other's payoff --- these ranges of $\alpha_i$ can be
interpreted as altruistic and spiteful, respectively.
Generalizing \citet{bester1998altruism},
\citet{possajennikov2000evolutionary} showed
that the unique ESS in this strategy space is
$\alpha_i = \alpha_j = \frac{\kappa}{2-\kappa}$.
Thus, when $\kappa > 0$, this ESS corresponds to players
with altruistic
preferences,
and when
$\kappa < 0$, their
preferences are spiteful.
Players who follow the subjective Nash equilibrium
with respect to 
$V_i^{\alpha_i}$ given by the altruistic ESS both receive
a higher payoff than the equilibrium of $\game$,
while the payoffs of the spiteful ESS are both
lower.
Since the Pareto-efficient symmetric
subjective Nash equilibrium is
at $\alpha_i = \alpha_j = 1$,
this means that as $\kappa \nearrow 1$,
the ESS approaches efficiency.
Intuitively, these other-regarding preferences
are stable in \citet{possajennikov2000evolutionary}'s model
because they serve as
commitment devices that elicit favorable responses from
the other player \citep{blush, commitment}.
That is, each agent best-responds under the 
assumption that the other player will play 
rationally with respect to their
utility function,
and
as
utility functions
are selected based on
payoffs from the opponent's best response
to the action optimizing
those 
utility functions,
the population converges to some $\alpha$.

\section{Setup}
\label{sec:setup}

We now discuss the formal framework on which our results
are based.
Let $V_i^\alpha(a_1,a_2) = \pay_i(a_1,a_2) + \alpha \pay_j(a_1,a_2)$
as above.
We say that a preference parameter $\alpha$ is \textit{egoistic}
if $\alpha = 0$, and \textit{other-regarding} otherwise.
In our results
we will use the following assumptions,
which are satisfied by the externality game:
\begin{enumerate}
    \item For any $\alpha_i, \alpha_j$, 
    $\nash_i(\alpha_i,\alpha_j)$ is unique. \label{assumption_ne_unique}
    \item For any $b$
    and $\alpha$,
    the function
    $h_i(a) = V_i^\alpha(a,b)$
    has a unique global maximum,
    $\br_i(b;\alpha)$.
    (That is, the best response to some action under any subjective utility function
    is unique.) \label{assumption_max_unique}
    \item For any $\alpha$, the function $g_i(\beta) = \nash_i(\beta, \alpha)$
    is surjective on $\mathbb{R}$.\footnote{For the externality game, there is no $\beta$ such that $g_i(\beta) = -\frac{m}{\kappa(1+\alpha)}$. However, one can check directly that $\max_{\act_i} \pay_i(\act_i, \br_j(\act_i;\alpha)) = \pay_i(\nash_i(\alpha,\alpha),\nash_j(\alpha,\alpha))$, which is the only condition for which this assumption is necessary.} \label{assumption_surjective}
\end{enumerate}

We give some remarks on the typical
indirect evolutionary models before
presenting our generalized model.
Recall our claim that 
the
strategy space assumed by much of
the indirect evolutionary game theory
literature
is too restrictive,
due to the assumption that
agents always play the Nash equilibrium
of $\subjgame$.
Playing a Nash equilibrium in response
to the other player's $\alpha_j$ can be exploitable,
in the sense that a player $j$ can ``force'' another
rational agent to play an action that is more
favorable to player $j$
(see Section~\ref{sub:strategies} for an example).
A player may
avoid being exploited in this way
by committing to some action,
independent of opponents' preferences.
We will therefore enrich the strategy space
in
$\game$
to relax
this assumption (Section~\ref{sub:strategies}).

Standard indirect evolutionary game theory
also assumes players perfectly observe
each other's
payoff functions
and subjective 
utility
functions.
This premise has been
questioned in previous work, e.g., \citet{dynamic_evo, greenbeards}.
We keep this assumption
due to
findings by
\citet{myopiclearning} and \citet{rational_learning}
that,
if
players use Bayesian updating
in repeated interactions with
each other, under certain conditions
they converge to accurate beliefs
about each other's 
utility functions
and play the Nash equilibrium.
\citet{evopref} and \citet{darwinnash} give
similar justifications for this assumption in their indirect evolutionary
models.

\subsection{Strategy space}
\label{sub:strategies}

Our strategy space combines the ``direct'' and ``indirect''
approaches to evolutionary game theory \citep{indirect}.
That is, this space includes both fixed actions of the base game
and strategies that choose actions as a function of
the player's own
subjective utility function and the
other player's strategy.

First, a \textit{behavioral strategy} plays an action $\act_i$,
independent of the other player's strategy.
The action $\act_i$ is common knowledge to both players
before $\game$ is played.
Second, as in the standard indirect evolutionary approach
\citep{bester1998altruism,possajennikov2000evolutionary},
a \textit{rational strategy} has 
a commonly known
preference parameter $\alpha_i$,
and 
always plays a best response
given $\alpha_i$
to their beliefs
about the other player.
A rational player believes that
another rational player
plays the Nash equilibrium of $\subjgame$.
Thus the best response to another rational player
with parameter $\alpha_j$ is $\nash_i(\alpha_i,\alpha_j)$.
A rational
player $i$ believes behavioral player $j$
plays
action~$\act_j$,
so the rational strategy is $\br_i(\act_j;\alpha_i)$.

To see the 
reason
for including both classes of strategies
in one model,
consider the externality game with $\kappa < 0$.
If a rational player~$i$ faces rational player~$j$ with
$\alpha_j =$~$\ 0$, and $\alpha_i = \alpha^* := \frac{\kappa}{2-\kappa} <$~$\ 0$,
we can check that
the payoff of $i$ increases while that
of $j$ decreases:
$\pay_i(\nash_i(\alpha^*,0), \nash_j(\alpha^*,0)) > \pay_i(\nash_i(0,0), \nash_j(0,0)) > \pay_j(\nash_i(\alpha^*,0), \nash_j(\alpha^*,0))$.
That is, $i$ can exploit the rationality of
$j$ by adopting an other-regarding preference parameter
as a commitment.
We therefore ask
what strategies are selected
for when we allow players to
\textit{ignore}
each other's commitments
(preferences),
in order to avoid exploitation,
and instead play some fitness-maximizing action.

In summary, our strategy space $\stratspace$ is the union of these sets:
\begin{enumerate}
    \item $\behavioral = \{B(\act) \ | \ \act \in \actspace\}$: Behavioral strategy whose
    action is $\respi(B(\act),\sigma_j) = \act$ for all $\sigma_j$.
    \item $\rational = \{R(\alpha) \ | \ \alpha \in \mathbb{R}\}$:
    Rational strategy whose action is
    $\respi(R(\alpha), \sigma_j) = \br_i(\act;\alpha)$ if
    $\sigma_j = B(\act)$,
    or
    $\respi(R(\alpha), \sigma_j) = \nash_i(\alpha,\alpha')$
    if
    $\sigma_j = R(\alpha')$.
\end{enumerate}

\section{Results}

We now characterize the Nash equilibria and stable strategies of $\stratspace$.
We show that
there are multiple neutrally stable strategies,
one of which acts according to egoistic preferences,
and no evolutionarily stable strategies.
This is in contrast to the results of
\citet{bester1998altruism} and \citet{possajennikov2000evolutionary},
who showed that without behavioral strategies,
a population with other-regarding preferences is the
unique ESS in
the externality game.
All proofs are in Appendix
\ref{app:proofs}.

\begin{proposition}
\label{nash_eqs_unpenalized}
Let $\game$ be a symmetric two-player game that satisfies assumptions \ref{assumption_ne_unique}- \ref{assumption_surjective}.
Then a strategy is a Nash equilibrium in $\stratspace$ if and only if it is either $B(\nash_i(0,0))$ or a strategy $R(\alpha)$ that is a Nash equilibrium in $\rational$.
Further, $B(\nash_i(0,0))$ is an NSS in 
$\stratspace$, and
$R(\alpha)$ is an NSS in $\stratspace$ if
and only if it is an NSS in $\rational$.
There are no ESSes.
\end{proposition}

Informally, a population that always plays the base game
Nash equilibrium can be invaded by rational players
with egoistic preferences, whose fitness against each
other matches that of the original population.
When the population consists of rational players
with other-regarding preferences that are stable
against other rational strategies, it can be
invaded by agents
that always play the Nash equilibrium of the
game with payoffs given by those same other-regarding
preferences.

\subsection{Complexity penalties}
\label{sub:complexity}

\paragraph{Single game}
Proposition \ref{nash_eqs_unpenalized} showed that strategies with either
egoistic or other-regarding preferences can be
neutrally stable,
and neither are
evolutionarily stable.
This
suggests 
that the standard indirect
evolutionary approach is insufficient to explain
the unique stability of other-regarding preferences.
However, our analysis above assumed that
players can use arbitrarily complex strategies
at no greater cost than simpler ones;
fitness is a function only of the
payoffs of strategies, not of the cognitive
resources required to use them \citep{richer_evo}.

We introduce complexity costs as follows.
For some complexity function $\complexfunc:\Sigma \to \mathbb{R}$,
we apply the usual evolutionary stability analysis
to a modified strategy fitness function:
\begin{align*}
    \fit_i(\sigma_i,\sigma_j) &= \pay_i(\respi(\sigma_i,\sigma_j),\respj(\sigma_j,\sigma_i)) - \complexfunc(\sigma_i).
\end{align*}

While behavioral strategies always play a fixed
action, rational strategies compute a best response
to each given opponent.
Within a single game, a behavioral strategy
thus requires less computation than a rational
strategy
(this assumption was also 
used by \citet{abreu2003evolutionary}).
Given this observation, for some
$\epsilon_R > 0$
we define $\complexfunc(\sigma) = \epsilon_R \mathbb{I}[\sigma \in \rational]$
(where the function $\mathbb{I}$ returns 1
if the condition in brackets is true,
and 0 otherwise).
Once this cost is accounted for, selection
favors the behavioral
strategy that plays the Nash equilibrium
of $\game$
(even when assumption \ref{assumption_surjective} does not hold).

\begin{proposition}
\label{simple_penalized}
Let $\game$ be a symmetric two-player game that satisfies assumptions \ref{assumption_ne_unique} and \ref{assumption_max_unique}.
Then for any $\epsilon_R >$~$0$,
the unique Nash equilibrium in $\stratspace$ under penalties
is $B(\nash_i(0,0))$, and this strategy
is an ESS.
\end{proposition}

An arbitrarily small cost of complexity
prevents rational strategies from matching the
fitness of the Nash equilibrium behavioral strategy.

\paragraph{Multiple games}

Proposition
\ref{simple_penalized},
again, appears inconsistent
with the stability of other-regarding preferences.
However, this result
is based on a metric of complexity that only accounts
for costs within one game --- the
cost of rational optimization versus playing
a constant action for any opponent ---
rather than cumulative costs \textit{across} games.
As \citet{waysofaltruism}
discuss qualitatively,
although agents who rely on situation-specific
heuristics avoid the fixed cost of explicit optimization paid by rational agents,
they do worse in some variable environments than the latter,
who can profit from having a general and
compact strategy of optimizing 
utility functions.
We formalize this tradeoff in this section.

Suppose that in 
each generation, the players in a population face a collection
of games $\{\game_1,\dots,\game_K\}$.
Each player uses a strategy that
(through the function $\respi$)
outputs an action conditional on both the
other player's strategy \textit{and}
the identity of the game.
One can apply the usual evolutionary stability
analysis to strategies that play the collection
of games,
by defining fitness as the sum of fitness
from each game
minus a multi-game complexity function
$\multicomplex$.
If a given strategy has $N(K)$ parameters
under selection across $K$ games,
$\multicomplex$ should
increase with $N(K)$.
An ideal definition of this function
would be informed by an accurate model of the
energetic costs of different kinds of cognition,
which is beyond the scope of this work.
We can define multi-game complexity
in our setting by generalizing the strategy
space from Section \ref{sec:setup}
to multiple games:
\begin{enumerate}
    \item $\multbehavioral = \{(B(a_k))_{k=1}^K\}$: Plays $a_k$ in game $\game_k$.
    \item $\multrational = \{R(\alpha)\}$: Plays the rational strategy
    with respect to $\alpha$ for each $\game_k$.
\end{enumerate}

The motivation for 
parameterizing a strategy in $\multrational$
by a single $\alpha$
is that, across a distribution of relevantly similar games
(e.g., variants of the externality game with different values of $\kappa$),
a rational player might be able to perform
well by 
interpolating
its other-regarding
preferences.\footnote{Compare to \citet{reciprocity_indirect}'s model
in which players evolve preferences for reciprocity
that they apply to both the ultimatum and dictator games.}
Then, for some $\epsilon_P > 0$,
letting $|X|$ denote the number of
unique elements
of $X$,
define:
\begin{align*}
    \multicomplex(\sigma) &= \begin{cases}
    \epsilon_P|\{a_k\}_{k=1}^K|, & \sigma \in \multbehavioral \\
    \epsilon_R + \epsilon_P, & \sigma \in \multrational.
    \end{cases}
\end{align*}

The set of stable strategies
under these multi-game penalties is
sensitive to the values of $\epsilon_R$ and $\epsilon_P$.
Intuitively,
a
behavioral
strategy will
be stable when $\epsilon_P$ is small, relative to
the profits this strategy can make by
adapting its response precisely to each game.
Conversely, when $\epsilon_P$ is sufficiently large, a rational strategy
can compensate for applying the same decision rule
to every game by avoiding the costs of game-specific heuristics.
In
the next section,
we show these patterns numerically.

\subsection{Evolutionary simulations with multi-game complexity penalties}
\label{sec:numerical}

Here, we will use an evolutionary simulation
algorithm to see how complexity costs across
games influence stable strategies --- in
particular, which (if any) other-regarding
preferences are selected?
For
simplicity,
we consider a set of just two
externality games
for a fixed $m$
with $\kappa = \kappa_1$ and $\kappa = \kappa_2$,
denoted $\oligone$ and $\oligtwo$.
However, to investigate the effects
of imbalanced environments (i.e.,
where $\oligone$ is played more or less frequently than $\oligtwo$)
we suppose that players spend a fraction $\propnegative$ of their time in
game $\oligone$ and $1-\propnegative$ in $\oligtwo$. 
Then, 
with $\pay_i^\kappa$ as the externality game payoff function for a given $\kappa$,
the multi-game penalized fitness of a strategy $\sigma$ against $\sigma'$
is:
\begin{align*}
    & \fit_i^{\kappa_1,\kappa_2}(\sigma,\sigma') = p\pay_i^{\kappa_1}(\respi(\sigma,\sigma',\oligone), \respj(\sigma',\sigma,\oligone)) \\
    &\qquad \negthinspace \negthinspace \negthinspace \negthinspace \negthinspace + (1-p)\pay_i^{\kappa_2}(\respi(\sigma,\sigma',\oligtwo), \respj(\sigma',\sigma,\oligtwo)) - c_2(\sigma).
\end{align*}

Due to the continuous strategy
space,
a replicator dynamic simulation is intractable.
Instead, we simulate an evolutionary process on $\stratspace$ using the \textit{adaptive learning}
algorithm \citep{adapt_learn},
implemented as follows
(details are in
Appendix \ref{app:expts}).
An initial population of size
$N=10$ is randomly sampled from the
spaces of rational and behavioral strategies.
In each round $t = 1,\dots,30$ of evolution,
each player in the population
either (with low probability) switches to a random strategy,
or else switches to the best response to a uniformly sampled
opponent in the population
(with respect to the penalized fitness
$\fit_i^{\kappa_1,\kappa_2}$ above).\footnote{This algorithm
is most appropriate when the evolutionary process is interpreted
as agents learning over their lifetimes,
updating their responses to each other,
rather than as genetic transmission.}
Note that a best response in the space $\multbehavioral$
might use one action across both games,
incurring a complexity cost of $\epsilon_P$
instead of $2\epsilon_P$.
We fix $\epsilon_R = 10^{-5}$ 
and $m = 1$.
In each experiment, we tune the multi-game complexity penalty $\epsilon_P$
(hereafter, ``per-parameter penalty'')
to approximately the smallest value necessary
to ensure that the population
almost
always converges to an element of $\multrational$
(a rational strategy).



\begin{figure}[H]
    \centering
    \includegraphics[width=7.5cm]{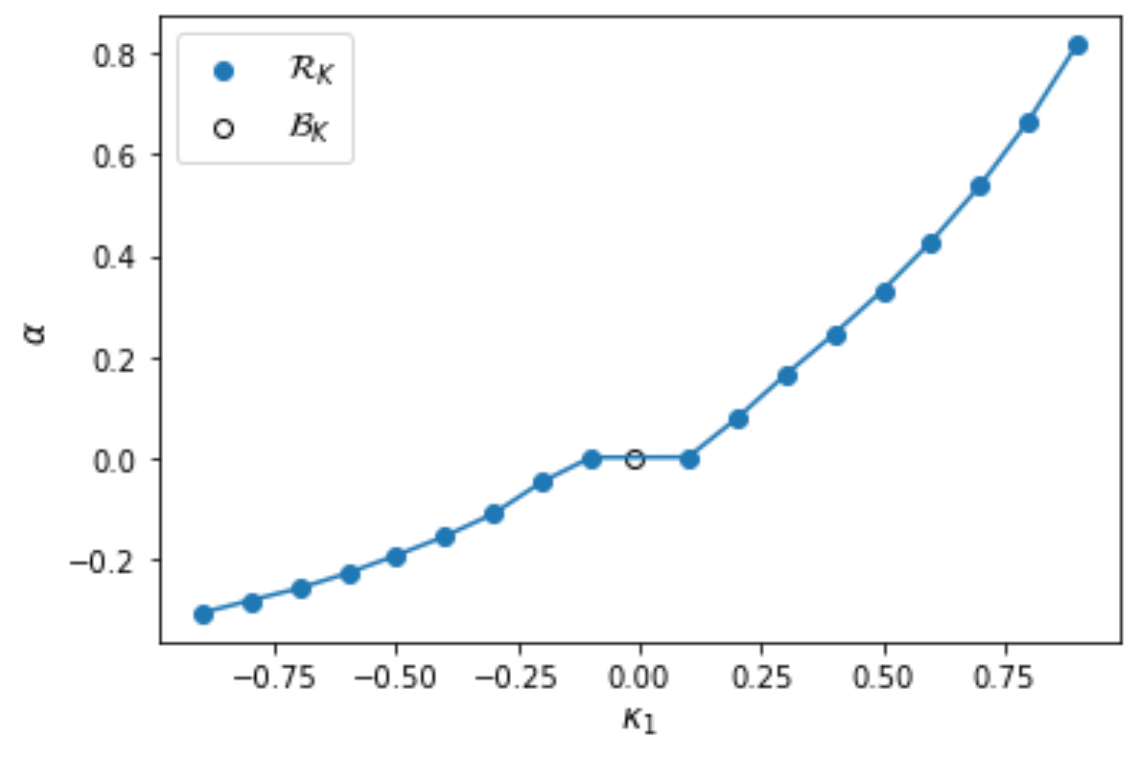}
    \caption{$\alpha$ values of the limit of an evolutionary simulation, 
    in which a proportion $0.5$ of games have $\kappa_1$ and a proportion
    $0.5$ have $\kappa_2=0.001$,
    as a function of $\kappa_1$.
    All members of the final population are in $\multrational$ except
    when $\kappa_1 = -0.01$, where the population consists of behavioral strategies (depicted with an open circle). The per-parameter penalty is $\epsilon_P = 0.001$.}
    \label{fig:vary_kappa_50_50}
\end{figure}

\paragraph{Varying strength of negative or positive externalities in one game}
First, we show that 
other-regarding preferences evolve under sufficiently strong 
negative or positive externalities,
given a sufficiently high per-parameter penalty.
We fix $\kappa_2 = 0.001$, $\propnegative = 0.5$, and $\epsilon_P = 0.001$,
and vary $\kappa_1 \in \{-0.9, -0.8, \dots, -0.1, -0.01, 0.1, \dots, 0.9\}$.
For $\kappa_1 = -0.01$,
the population converged to a behavioral
strategy that uses only one action, for all values of
$\epsilon_P$ we tested
(see the open circle in Figure~\ref{fig:vary_kappa_50_50}).
This suggests that when both games
are sufficiently similar,
a behavioral strategy can 
interpolate
across
both games at less expense
than a rational strategy.
Figure~\ref{fig:vary_kappa_50_50} shows that, as expected,
the sign and magnitude of the stable~$\alpha$ value
scales with $\kappa_1$.
For $\kappa_1 \in [-0.1, 0.1]$,
the population converges to $\alpha \approx 0$, suggesting
that other-regarding preferences only
interpolate
well across
these externality games when the externalities are sufficiently
strong in magnitude.

\begin{figure}[H]
    \centering
    \includegraphics[width=7.5cm]{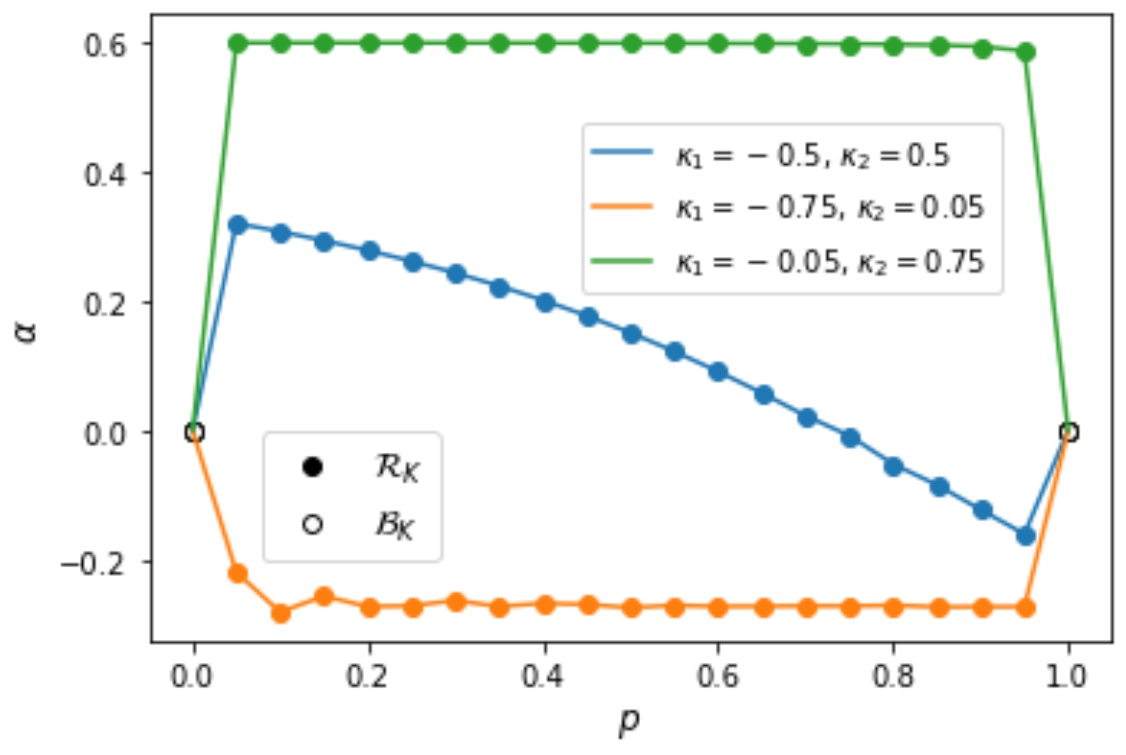}
    \caption{$\alpha$ values of the limit of an evolutionary simulation,
    in which a proportion $p$ of games have $\kappa_1$ and a proportion
    $1-p$ have $\kappa_2$,
    as a function of $\propnegative$, with three pairs of~$\kappa$ values.
    All populations are in $\multrational$ except $\propnegative = 0.00005$ and $\propnegative = 0.99995$, where the population consists of behavioral strategies (depicted with open circles). The per-parameter penalty is $\epsilon_P = 0.002$.}
    \label{fig:vary_prop_-0,5_0,5}
\end{figure}

\paragraph{Varying proportion of games with negative versus positive externalities}
Next, we show that the strength of altruism versus spite in the limiting population
scales nonlinearly with the proportion of games
with negative versus positive externalities.
With $\epsilon_P = 0.002$,
we vary the fraction of games with $\kappa <$~$ \ 0$,
over $\propnegative \in \{0.00005, 0.05, 0.1, \dots, 0.9, 0.95, 0.99995\}$,
for three pairs of games.
For all pairs of $\kappa$ in this
experiment, the values $\propnegative = 0.00005$ and $\propnegative = 0.99995$ have 
one-action behavioral strategies in
the limiting population
(see the open circles in Figure \ref{fig:vary_prop_-0,5_0,5}).
When one game is extremely rare,
the rational strategy's gains from
interpolation
across games do not outweigh
the cost $\epsilon_R$ of rationality.

First we fix $\kappa_1 = -0.5$ and $\kappa_2 = 0.5$
(blue curve in Figure \ref{fig:vary_prop_-0,5_0,5}).
Again, the trend of decreasing $\alpha$ with greater~$\propnegative$ is as expected,
though there is a bias towards altruism:
an equal proportion of positive and negative
externalities gives $\alpha >$~$0$.
When $\kappa_1 = -0.75$ and $\kappa_2 = 0.05$ (orange curve),
even small proportions of the large-magnitude
negative $\kappa_1$ are sufficient for the rational
population to adopt $\alpha < 0$,
and $\alpha$ remains roughly constant
above $\propnegative \approx 0.1$.
That is, in an environment where one game has weak positive externalities and the other has strong negative externalities,
most of the effect
on the population's
other-regarding preferences comes just from having 
a frequency of strong negative externalities \textit{above some
(small) threshold}.
The same pattern holds in the opposite direction when $\kappa_1 = -0.05$ and $\kappa_2 = 0.75$ (green curve).

\begin{figure}[H]
    \centering
    \includegraphics[width=7.5cm]{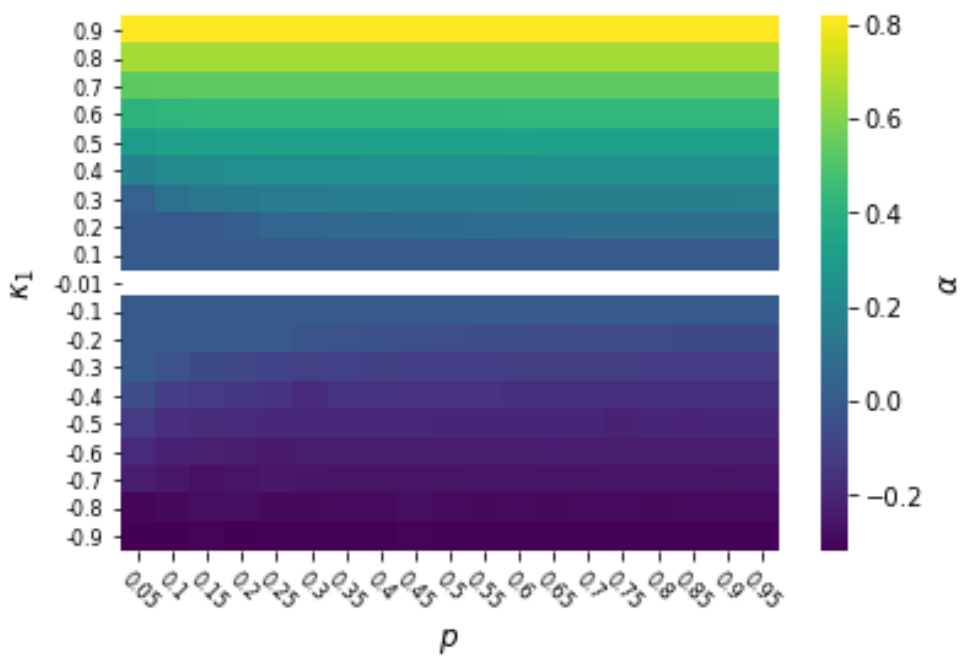}
    \caption{$\alpha$ values of the limit of an evolutionary simulation
    in which a proportion $p$ of games have $\kappa_1$ and a proportion
    $1-p$ have $\kappa_2=0.001$, as a function of $\kappa_1$ and $\propnegative$. White cells indicate that the limiting population is not in $\multrational$.
    The per-parameter penalty is $\epsilon_P = 0.002$.}
    \label{fig:heatmap}
\end{figure}

In Figure \ref{fig:heatmap}, we vary 
both $\kappa_1$ and $\propnegative$,
keeping $\kappa_2 = 0.001$.
For any $\propnegative$, the result
from Figure \ref{fig:vary_kappa_50_50}
where a rational strategy is not stable
for small $\kappa_1$ still holds.
Likewise, the result that $R(0)$ takes over
the population when $\kappa \in [-0.1, 0.1]$
is not sensitive to $\propnegative$.
Generalizing the trend from Figure
\ref{fig:vary_prop_-0,5_0,5},
for sufficiently large magnitudes of
$\kappa_1$,
only a minority of games need to have
$\kappa$ far from 0
for strong other-regarding
preferences to be stable.



\begin{figure}[H]
    \centering
    \includegraphics[width=7.5cm]{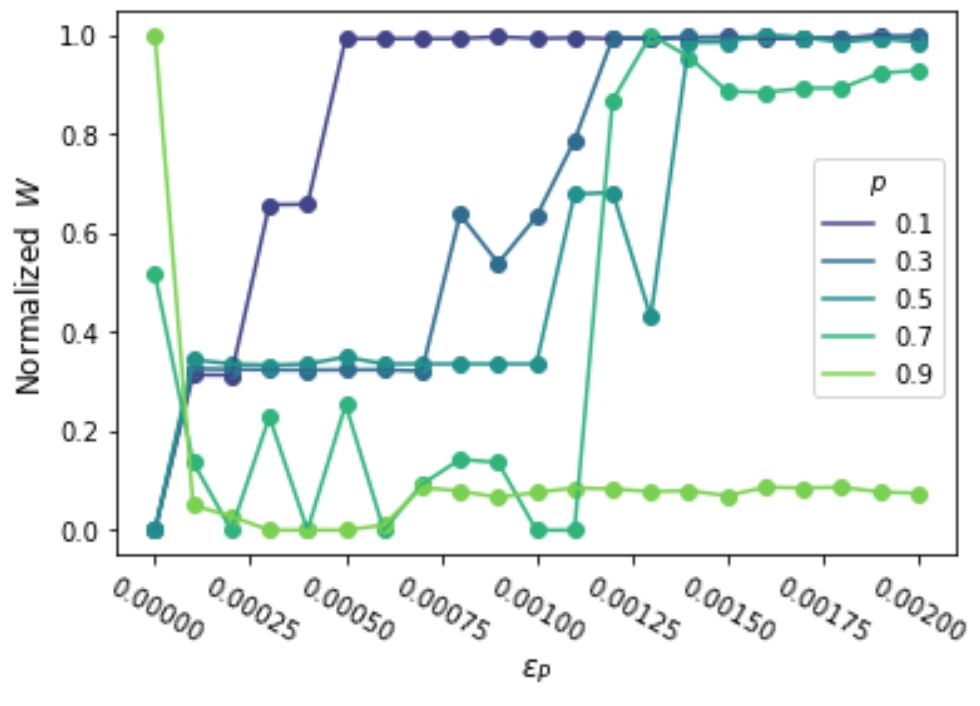}
    \caption{Normalized social welfare of the limit of an evolutionary simulation,
     in which a proportion $\propnegative$ of games have $\kappa_1=0.5$ and a proportion
    $1-\propnegative$ have $\kappa_2=-0.5$, as a function of the per-parameter penalty $\epsilon_P$, for different values of~$\propnegative$.} 
    \label{fig:welfare_vary_pen}
\end{figure}

\paragraph{Social welfare in the limiting population as a function of the 
per-parameter penalty}
Finally, we show how the total payoffs of the limiting population vary both
with the 
size of the per-parameter penalty,
and with the proportion of games with positive versus negative externalities.
Fixing $\kappa_1 = -0.5$ and $\kappa_2 = 0.5$,
we vary
$\epsilon_P \in \{0, 0.0001, \dots, 0.0019, 0.002\}$
for each $\propnegative \in \{0.1, 0.3, 0.5, 0.7, 0.9\}$.
To visualize the transitions
between limiting populations
of behavioral versus
rational strategies,
we compute the social welfare $W_{\epsilon_P,\propnegative} = \pay_i + \pay_j$ 
averaged over
the last two rounds (for some parameter values, the population oscillates)
of each evolutionary
simulation for penalty $\epsilon_P$
and proportion $\propnegative$, shown in Figure \ref{fig:welfare_vary_pen}.
\footnote{
We take the average $\overline{W}_{\epsilon_P,\propnegative}$ over
10
runs of each simulation,
and given the list $\mathcal{W}_\propnegative = \{\overline{W}_{0,\propnegative}, \dots \overline{W}_{0.00225,\propnegative}\}$,
we normalize each $\tilde{W}_{\epsilon_P,\propnegative} = \frac{\overline{W}_{\epsilon_P,\propnegative} - \min \mathcal{W}_\propnegative}{\max \mathcal{W}_\propnegative - \min \mathcal{W}_\propnegative}$.
(This is for ease of visualization.)
Note that the trend in Figure \ref{fig:welfare_vary_pen} for $\propnegative = 0.7$ is
exaggerated by normalization; the stable rational value
is $\alpha = 0.0004$, so the social welfare
does not actually vary significantly.
}

For
most values of $\propnegative$,
when
there is no 
per-parameter
penalty
($\epsilon_P = 0$)
the population attains
the near-lowest social welfare, where
all in the population
play the 
base game Nash equilibrium.
The penalty $\epsilon_P = 0.0015$ is sufficient
for all populations to converge to an
other-regarding rational strategy, which
attains the highest social welfare when $\propnegative \leq 0.7$
but 
nearly the lowest
when $\propnegative = 0.9$, i.e., when most of the games have $\kappa < 0$.
For intermediate values of $\epsilon_P$,
the population oscillates between $R(0)$
and a behavioral
best response to $R(0)$ in each 
game, usually resulting
in social welfare 
between that of very low or high $\epsilon_P$.
The minimum value of $\epsilon_P$
necessary for convergence to the rational
strategy is largest for values of $\propnegative$
closest to 0.5, while only a small penalty is necessary
when $\propnegative = 0.1$ or~$0.9$
(see the values of $\epsilon_P$ where
the curves in Figure \ref{fig:welfare_vary_pen} plateau).
Intuitively,
if the large majority
of games have the same $\kappa$, a behavioral
strategy does not profit much from adapting with multiple
actions, relative to the complexity costs 
of playing different actions for two games.


The magnitude of $\epsilon_P$ relative to $\epsilon_R$
required for other-regarding preferences to be stable might appear unrealistically large, based on these results.
We note
the distinction between the 
fixed cognitive costs of
developing a rational decision procedure,
and the per-use costs of learning heuristics for each context
and recognizing when each is appropriate.
\citet{finite_complexity} argues that lexicographic, or infinitesimal, complexity costs are appropriate for the 
former --- these start up costs are a tiebreaker between strategies
that are otherwise equally capable ---
while finite non-negligible costs are suitable for the latter.
It is therefore plausible that in several evolutionary contexts,
the costs of adapting to each interaction
from scratch
outweigh costs of rationality.
Regardless, given the sensitivity of the stable populations in these
experiments to $\epsilon_P$,
it is important to account for the relative strength
of these two factors when predicting the result of an evolutionary process.

\subsection{Inefficiency and the secret handshake}

Lastly, we
discuss the implications of complexity costs
for
another model that appears to preclude
the evolution of certain other-regarding preferences.
Recall that we have defined the
utility functions
of rational strategies as the player's own payoff
plus a multiple of the opponent's payoff.
Previous work has shown (in finite-action games)
that if \textit{all} possible 
subjective utility
functions are
permitted,
and players observe each other's subjective
utility functions,
then all stable strategies achieve a Pareto 
efficient payoff \citep{evopref,darwinnash}.
This conclusion follows from the ``secret handshake'' argument:
a player who is indifferent among all action pairs can
select an equilibrium that matches any other strategy's
action against that strategy,
but plays an action achieving an efficient
payoff against itself \citep{handshake}.
These results rule out both the base game Nash equilibrium and the ESS in $\rational$ of the externality game,
which is $R(\alpha^*)$ for $\alpha^* = \frac{\kappa}{2-\kappa} < 1$,
while $R(1)$ is the unique
efficient rational strategy.

One might suspect, then, that
our conclusion from the numerical experiments ---
i.e., inefficient other-regarding preferences
can be stable when agents play multiple games ---
would not hold after including the strategy classes
from \citet{evopref} and \citet{darwinnash}.
When we include complexity costs,
however, the secret handshake argument does
not follow.
Let $\handshake$ be the class of strategies whose
subjective utility functions are constant over all
action pairs, and which use the equilibrium selection rule
described above.
Because this strategy requires
choosing different Nash equilibria depending on the opponent,
we claim that it is more complex than either a behavioral or rational
strategy.
For $\epsilon_H > \epsilon_R$,
let $\complexfunc(\sigma) = \epsilon_H\mathbb{I}[\sigma \in \handshake] + \epsilon_R \mathbb{I}[\sigma \in \rational]$.
Then $B(\nash_i(0,0))$ is still an ESS
under the conditions of Proposition \ref{simple_penalized},
with $\handshake$ added to the strategy space.
The proof is straightforward;
given a positive penalty,
a strategy from $\handshake$
cannot match the payoff of $B(\nash_i(0,0))$ against itself,
by the definition of the base game
Nash equilibrium:
\begin{align*}
    & \max_{\sigma \in \handshake} \fit_i(\sigma, B(\nash_i(0,0))) \\
    &\qquad = \pay_i(\nash_i(0,0), \nash_j(0,0)) - \epsilon_H \\
    &\qquad < \fit_i(B(\nash_i(0,0)),B(\nash_i(0,0))).
\end{align*}
We conjecture that across multiple games, a sufficiently
large penalty $\epsilon_H$ would yield similar
results to Section \ref{sec:numerical}.

\section{Discussion}

The puzzle that motivated this work was
the apparent prevalence of other-regarding
preferences, such as altruism and spite,
despite the
possibility of selection for commitment strategies
that ignore the signals of other-regarding preferences.
Our results suggest that this puzzle stems from
a neglect of complexity considerations
in previous literature on the evolution of preferences.
We considered a class of two-player symmetric
games that includes the games used by \citet{bester1998altruism} and \citet{possajennikov2000evolutionary}
to illustrate the stability of altruism and spite.
First, via evolutionary stability
analysis on a strategy space that combines
the direct and indirect approaches,
we confirmed
that other-regarding preferences are no longer
uniquely stable when fixed-action strategies
can also evolve.
We then showed numerically that, although other-regarding
preferences are unstable when agents play a single
game under costs of strategy complexity,
if the costs of distinct fixed actions across \textit{multiple} games are sufficiently high,
other-regarding preferences are stable.
These costs also explain why inefficient stable strategies
can persist --- the flexible ``secret handshake'' strategy,
which has been purported to guarantee that stability implies
efficiency,
is too complex to invade populations with certain inefficient strategies.

Accounting for the costs of adapting strategies to specific games
plausibly sheds light on other phenomena in evolutionary game theory.
For example, \citet{punish_multiplicity} argued that a common explanation of cooperation as a
product of punishment,
e.g., as in tit-for-tat in the repeated Prisoner's Dilemma,
proves too much:
``Moralistic'' strategies, which not only
punish noncooperation but also punish
those who do not punish noncooperation,
can enforce the stability of \textit{any}
individually rational behavior.
These moralistic strategies require sophisticated recognition of the behaviors
that constitute cooperation or punishment in each given game.
If some individually rational behavior
enforced by a moralistic strategy is only
marginally better for the cooperating player than getting punished,
another strategy could invade by avoiding
the complexity cost of the moralistic strategy,
which outweighs the direct fitness cost
of being punished.
Thus, under complexity costs,
the set of evolutionarily stable behaviors may be
much smaller.
It is also important to note that classes of simple, generalizable
utility functions
other than those we have considered
might evolve.
Instead of having
utility functions
given by their payoff plus a multiple of
the other agent's payoff,
agents could develop
utility functions
with
an aversion to exploitation or inequity \citep{HUCK199913,ineq_aversion}.
Future work could investigate selection pressures
on utility functions of different complexity.

Besides explaining biological behavior,
our model of complexity-penalized preference
evolution might also motivate predictions
of the behavior of artificial agents, such as reinforcement learning (RL) algorithms.
Policies are updated based on reward signals
similarly to fitness-based
updating of populations in evolutionary models
\citep{rl_replicator}.
It is common in RL training to penalize
strategies
(``policies'')
according to their complexity,
and deep learning researchers
have argued that artificial neural
networks have an implicit bias towards
simple functions
\citep{sgd_bayesian, deep_generalizes}.
Thus,
RL agents
trained together 
may
develop other-regarding preferences, as far as
the assumptions of our model are satisfied by
the tasks these agents are trained in.
A better understanding of the relationship between complexity costs and the distribution of environments these agents are trained in may help us better understand what kinds of preferences they acquire.

\bibliography{refs}
\end{multicols}

\newpage

\appendix

\section{Proofs}
\label{app:proofs}

\subsection{Proof of Proposition \ref{nash_eqs_unpenalized}}


\paragraph{Behavioral}
Define $\nashpay = \pay_i(\nash_i(0,0), \nash_j(0,0))$, the payoff of the
Nash equilibrium with egoistic preferences.
By the definition of Nash equilibrium of $\game$,
since $\max_{a_i} \pay_i(a_i, \nash_i(0,0)) = \nashpay$, the strategy $B(\nash_i(0,0))$ is
a Nash equilibrium in $\stratspace$.
Suppose $a \neq \nash_i(0,0)$.
Then
we must have $\max_{a_i} \pay_i(a_i, a) > \pay_i(a,a)$, because otherwise
uniqueness of the Nash equilibrium (assumption \ref{assumption_ne_unique}) would be violated.
So $B(a)$ is not a Nash equilibrium in $\stratspace$.

Since the Nash equilibrium of $\game(0,0)$ is unique, 
there is no behavioral strategy $B(a) \neq B(\nash_i(0,0))$ such that 
$\fit_i(B(a), B(\nash_i(0,0))) = \fit_i(B(\nash_i(0,0)), B(\nash_i(0,0)))$.
Suppose a rational strategy $R(\alpha)$ satisfies
$\fit_i(R(\alpha), B(\nash_i(0,0))) = \fit_i(B(\nash_i(0,0)), B(\nash_i(0,0)))$.
Then $\argmax_{a} V^\alpha_i(a,\nash_i(0,0)) = \argmax_{a} \{\pay_i(a,\nash_i(0,0)) + \alpha \pay_j(a,\nash_i(0,0))\} = \nash_i(0,0)$.
(This is satisfied for $\alpha = 0$.) But this implies that $\nash_i(0,0) = \nash_i(\alpha,\alpha)$.
So
$\fit_i(R(\alpha),R(\alpha)) = \pay_i(\nash_i(\alpha,\alpha),\nash_j(\alpha,\alpha)) = \nashpay$,
and $\fit_i(B(\nash_i(0,0)), R(\alpha)) = \nashpay$,
therefore $B(\nash_i(0,0))$ is neutrally stable
(but not an ESS).

\paragraph{Rational} 
It is immediate that $R(\alpha)$ can only be a Nash equilibrium
in $\stratspace$ if it is a Nash
equilibrium in $\rational$.
Let  $R(\alpha)$ be such a strategy.
$R(\alpha)$
always plays $\nash_i(\alpha, \alpha)$ against itself,
so $\fit_i(R(\alpha), R(\alpha)) = \pay_i(\nash_i(\alpha, \alpha), \nash_j(\alpha, \alpha))$.
Suppose a deviator $\sigma$ plays
$a_i$.
Given 
assumption \ref{assumption_surjective},
for any $a_i$ there exists a $\beta$
such that $a_i = \nash_i(\beta, \alpha)$.
Therefore:
\begin{align*}
    \fit_i(\sigma, R(\alpha)) &\leq \max_{a_i} \pay_i(a_i, \br_j(a_i;\alpha)) \\
    &= \max_{\beta} \pay_i(\nash_i(\beta, \alpha), \br_j(\nash_i(\beta, \alpha);\alpha)) \\
    &= \max_{\beta} \pay_i(\nash_i(\beta, \alpha), \nash_j(\beta, \alpha)) \\
    &= \pay_i(\nash_i(\alpha, \alpha), \nash_j(\alpha, \alpha)).
\end{align*}
Where the last line follows because $R(\alpha)$
is a Nash equilibrium in the space of rational strategies.
So $R(\alpha)$
is a Nash equilibrium in $\stratspace$.

Suppose $R(\alpha)$ is neutrally stable in $\rational$.
By assumption \ref{assumption_max_unique}, $\nash_i(\alpha, \alpha)$ is the unique
action $a$ such that $\fit_i(B(a), R(\alpha)) = \fit_i(R(\alpha),R(\alpha))$.
Then $\fit_i(B(\nash_i(\alpha,\alpha)), R(\alpha)) = \pay_i(\nash_i(\alpha,\alpha), \nash_j(\alpha,\alpha)) = \fit_i(R(\alpha),R(\alpha))$,
and $\fit_i(B(\nash_i(\alpha,\alpha)), B(\nash_i(\alpha,\alpha))) = \pay_i(\nash_i(\alpha,\alpha), \nash_j(\alpha,\alpha)) = \fit_i(R(\alpha), B(\nash_i(\alpha,\alpha)))$.
So $R(\alpha)$ is neutrally stable (but not an ESS). 
On the other hand, if $R(\alpha)$ is not an NSS in $\rational$,
the same counterexample to neutral stability applies
in the expanded space $\stratspace$,
thus $R(\alpha)$ is not an NSS in $\stratspace$.

\subsection{Proof of Proposition \ref{simple_penalized}}

\paragraph{Behavioral}
The conditions for Nash equilibrium in $\behavioral$
do not change, since this set has the lowest complexity.
However, when assessing the stability of $B(\nash_i(0,0))$,
it suffices to only consider invader strategies in $\behavioral$,
because for a strategy $\sigma \in \rational$, $\fit_i(\sigma, B(\nash_i(0,0))) \leq \nashpay - \epsilon_R < \fit_i(B(\nash_i(0,0)), B(\nash_i(0,0)))$.
Since the Nash equilibrium of $\game(0,0)$ is unique (assumption \ref{assumption_ne_unique}), 
there is no other behavioral strategy $B(a)$ such that 
$\fit_i(B(a), B(\nash_i(0,0))) = \fit_i(B(\nash_i(0,0)), B(\nash_i(0,0)))$.
Thus $B(\nash_i(0,0))$ is an ESS under penalties.

\paragraph{Rational}
Let $R(\alpha)$ be any rational strategy. Then $\fit_i(R(\alpha), R(\alpha)) = \pay_i(\nash_i(\alpha, \alpha), \nash_j(\alpha, \alpha)) - \epsilon_R$.
But $\fit_i(B(\nash_i(\alpha,\alpha)), R(\alpha)) = \pay_i(\nash_i(\alpha, \alpha), \nash_j(\alpha, \alpha))$,
so $R(\alpha)$ cannot be a Nash equilibrium under penalties.

\section{Details on Numerical Experiments}
\label{app:expts}

Each strategy is parameterized by $(\alpha, \act_1, \act_2, \stratnum)$,
where the strategy is
$R(\alpha)$ if $\stratnum = 0$
or 
$(B(\act_1),B(\act_2))$ if $\stratnum = 1$. 
A population of size $N = 10$ is initialized with $\alpha, \act_1, \act_2 \stackrel{i.i.d.}{\sim} N(0,1)$
and
$n \stackrel{i.i.d.}{\sim} \text{Bern}(0.5)$
for each player in the population.
Let $q_1 = 0.01$ and $q_{t+1} = q_t \cdot \frac{t}{t+1}$
if $t \leq 20$,
otherwise $q_{t+1} = 0$.
The probability of switching to a random strategy
from the initialization distribution in round $t$
of evolution is $q_t$.
(We decay $q_t$ to decrease the rate of stochasticity
and thus help convergence.)
Best responses
in the space of $\multbehavioral$ 
are computed analytically; for
$\multrational$, we use gradient ascent.


\end{document}